\documentclass[showpacs,preprintnumbers,amsmath,amssymb,epsf]{revtex4}
\usepackage{graphicx,epsfig}
\usepackage{bm}
\newcommand{\be}{\begin{equation}}
\newcommand{\ee}{\end{equation}}
\newcommand{\bse}{\begin{subequations}}
\newcommand{\ese}{\end{subequations}}
\newcommand{\bary}{\begin{eqnarray}}
\newcommand{\eary}{\end{eqnarray}}

\begin{document}


\title{Very High Energy Cosmic Rays from Centaurus A}
\author{Nissim Fraija$^{*}$, Sarira Sahu$^{*}$ and Bing Zhang$^{\dagger}$}
\affiliation{$^*$Instituto de Ciencias Nucleares,
Universidad Nacional Aut\'onoma de M\'exico, 
Circuito Exterior, C.U., A. Postal 70-543, 04510 Mexico DF, Mexico\\
$^{\dagger}$Department of Physics and Astronomy,
University of Nevada, Las Vegas, NV 89154, USA
}


\begin{abstract}

Centaurus A is the nearest radio-loud AGN and is detected from radio
to very high energy gamma-rays.  Its nuclear spectral energy
distribution shows two peaks, one in the far-infrared band and another
at about 150 keV. By assuming the second peak is due to the electron
synchrotron emission and the power index for the differential
spectrum of the very high energy cosmic ray proton to be 2.7 
we show that only $pp$ interaction is
responsible for the observed GeV-TeV emission from Centaurus A. 
We also found that indeed  many very high energy cosmic ray  protons from
Centaurus A can arrive on Earth thus supporting the recent observation
of two events by Pierre Auger Observatory. 

\end{abstract}

\maketitle

\section{Introduction}

Centaurus A (Cen A or NGC 5128) is the nearest active  radio galaxy
with a distance of approximately 3.8 Mpc and redshift 
$z=0.002$.  Because of its proximity, Cen A is one of the best
studied radio galaxies although its bolometric luminosity is 
not large by an AGN standard. Optically, Cen A is an elliptical galaxy
undergoing late stages of a merger event with a small 
spiral galaxy.   From Cen A, sufficiently large amount of photometric
data is available to build a well sampled spectral energy 
distribution (SED).  The emission from the nucleus of Cen A has been
observed in radio, infrared, X-ray, 
$\gamma$-ray\cite{Winkler,Mushotzky,Bowyer:1970,Baity:1981,Hardcastle:2003ye} 
and also in the 
GeV-TeV 
range\cite{Sreekumar:1999xw,Aharonian:2009xn,Chadwick:1999my,Kabuki:2007am,Allen:1993ep,Rowell:1999fx,Abdo:2009wu}.  Recently the Fermi Large Area
Telescope (LAT) has also observed Cen A in the energy range 0.2 to 100
GeV\cite{Abdo:2009wu}. Observations in different wavelengths  
show that Cen A has FR I morphology having 
two radio lobes, non-blazar source with a jet inclination of about
$50^{0}$.  It  has a central supermassive black 
hole of mass $m_{BH}\sim (0.5-1.2) \times 10^8\, M_{\odot}$.  The
nuclear SED shows two peaks, one  in the 
far-infrared band ($\sim 4\times 10^{-2} eV$)   and another at about 150 keV. 

High energy $\gamma$-Rays are produced due to non-thermal process
where the particles are accelerated to ultra relativistic energies.
Detection of GeV-TeV photons from Cen A signifies that, Cen A has the
potential to produce very high energy cosmic rays. Recently Pierre Auger Collaboration reported 
a correlation between ultra high energy cosmic rays (UHECRs) and nearby AGN within $\sim 75$ Mpc. 
In particular two of these events fall within $3.1^0$ around Cen A,
thus strengthening the possibility that this object is a strong
candidate for UHECR source. By assuming the two events are from Cen A
the expected rate of UHE neutrinos in detectors like IceCube are
estimated \cite{Halzen:2008vz,:2007qd}. By using the same hypothesis the diffuse neutrino
flux from Cen A are estimated\cite{Koers:2008hv}. Also the flux of
high energy cosmic rays  as well as the accompanying expected secondary 
photons and neutrinos are calculated from hadronic models\cite{Kachelriess:2008qx}.

It is logical to argue that the astrophysical objects which are
producing UHECRs also produce high energy $\gamma$-rays due to
interaction of the UHECRs with the 
background\cite{Kachelriess:2008qx,Romero:1995tn,Isola:2001ng,Honda:2009xd,Kachelriess:2009wn}. We have to
also keep in mind that apart from the hadronic interaction ($pp$ and
$p\gamma$) with the
background, there are leptonic processes (e.g. electron synchrotron,
Inverse Compton (IC) Scattering, etc.) also responsible for the production
of   high energy $\gamma$-rays and efficiencies of both the hadronic
and leptonic processes depend on the background particles and/or the
magnetic field.  By using the currently available GeV-TeV gamma-ray
data from Cen A,  it has been stressed in the literature that
only hadronic interaction is insufficient to explain the two events
from Cen A\cite{Gupta:2008tm}.

Our paper is organized as follows: In section 2, we discuss different
processes which are responsible for the production of gamma-ray in Cen
A and how it is compared with the observed photon spectrum. Section 3
is devoted to estimate the number of events from Pierre Auger
Observatory (PAO) and its relation with the CR normalization constant.
A summary of the observations of very high energy gamma-rays above 250 GeV by
different experiments and calculation of expected events from these
experiments are discussed in section 4. We briefly conclude our
results in section 5.

\section{Gamma-Ray Emission Process}

The main mechanisms responsible for production of $\gamma$-rays in
astrophysical objects are the: synchrotron emission by charge
particles (protons and electrons), 
inverse Compton (IC) scattering of electron by background photons,
inelastic scattering of high energy cosmic ray proton with the
background protons and the photons, and also the
photo-de-excitation of nuclei.  Depending
on the projectile energy and the nature of the surrounding environment of the astrophysical
object some or many of the above processes can be effective. Here we
study  the case of Cen A  where we have to analyze which are the most probable
mechanisms to produce the high energy $\gamma$-rays\cite{Ribicki,Aharonian:book}.

In general for mildly relativistic systems like AGNs, proton synchrotron
radiation is an inefficient process in  comparison to electron
synchrotron radiation and the energy loss rate of  proton is
$(m_p/m_e)^4\sim 10^{13}$ times slower than the electron. 
Also emitted
photon energy from the proton is $(m_p/m_e)^3\sim 6 \times 10^9$ times
smaller than the photon emitted by the electron of same energy as
proton.  The above analysis is also true for a relativistic
system\cite{Gupta:2007yb} where proton synchrotron process is
suppressed in comparison to the leptonic one.
So in this case we neglect the proton synchrotron radiation
process as compared to the electron synchrotron one.  On the contrary  in Gamma-Ray
Bursts, where the jet is ultra-relativistic, synchrotron emission by
proton can be an efficient process. Also the photo-de-excision
contribution to the $\gamma$-ray is very small in Cen A environment,
so we do not take this into account here.
In general, the energetic electrons which are
Fermi accelerated in the jet of Cen A can have a power spectrum given
by    
\be
\frac{dN_e}{dE_e} = A_e 
 \left\{ 
\begin{array}{l l}
 \left ( \frac{E_e}{E_0} \right )^{-\alpha} & \quad E_e < E_{e, b} \\
  \frac{E_{e,b}}{E_0} \left ( \frac{E_e}{E_0}\right )^{-(\alpha+1)}& \quad  E_e \ge E_{e,b} \\
\end{array} \right. .
\ee
The constant $A_e$ has the dimension of $photon/keV/cm^2/s$
and $E_0$ is the energy normalization constant which can vary
depending on the range of energy one considers.
The above electrons will lose energy through synchrotron emission and the
break in the electron energy  $E_{e,b}$ is   given by
\be
E_{e,b}= \frac{6 \pi m^2_e }{\sigma_T\xi_B}   \frac{\Gamma^5 }{ \beta^2}  f_{es} \frac{\Delta t^{obs}}{ L^{obs}_{\gamma}},
\ee
and correspondingly the break in the photon energy is given by
\be
E^{obs}_{\gamma, b} =\frac{27 \pi^3 e m_e}{2 \sigma^2_T}   
f^2_{es} \frac{\Gamma^8}{\beta^4} \left ( \frac{1}{2 \xi_B}   \right )^{1/2} \frac{\Delta t^{obs}}{( L^{obs}_{\gamma})^{3/2}},
\ee
where $\sigma_T \simeq 6.65 \times 10^{-25}\, cm^2$ is the Thompson
scattering cross section and $\beta=[(\Gamma^2-1)/\Gamma^2]^{1/2}$. Also $\Gamma$, $\xi_B$ , $\Delta t^{obs}$
and $f_{es}$ are respectively the bulk Lorentz factor within the jet,  
fraction of energy carried by the magnetic field, observed time
variability and ratio 
of shell expansion time to synchrotron emission time which is given by
\be
f_{es}(E_e)=
 \left\{ 
\begin{array}{l l}
 \frac{E_e}{E_{e,b} } & \quad  E_e < E_{e, b} \\
  1 & \quad E_e \ge E_{e,b}\\
\end{array} \right. .
\ee 
The $L^{obs}_{\gamma}$ corresponds to the observed luminosity in the
observed range of photon energies. 

With the synchrotron self-Compton model, M. Chiaberge et al.  have
reproduced the whole nuclear emission of Cen A 
by taking the bulk Lorentz factor $\Gamma \le 3- 5$ and these 
values are consistent with the  
mildly relativistic\cite{Orellana:2009nz, Chiaberge:2001ek} or a  little more
\cite{Lenain:2008ze}  
proper motions observed on sub-parsec scales\cite{Jones:1996si}.  
Also in ref.\cite{Meisenheimer:2007zv} using SSC model they estimate  
the $\Gamma < 2.5$ in parsec scale.
 The observation of Cen A during 1973 to 1983 in X-rays below 100 keV
 has shown greater than an order of magnitude variation on time scales
of years \cite{Bond:1996,Turner:1997,Israel:1998ws}. Later
observations 
have also shown variability on similar 
timescales \cite{Grandi:2003eb,Rothschild:2005jp} 
but no high luminosity state was detected since 1985.

In the present work we assume that
the second peak at 150 keV in the nuclear SED region is  solely due to the
electron synchrotron emission in the nuclear region of Cen A and we  use the energy
normalization $E_0=100$ keV \cite{Jourdain:1993}.  The free
parameters in the electron synchrotron model are $\Gamma$, $\xi_B$ ,
$\Delta t^{obs}$ and $f_{es}$.  But from the above discussions we know
that jet in Cen A is mildly relativistic and around 100 keV range the
time variability is of order years. In order to fix the observed photon
energy $E^{obs}_{\gamma, b} =150$ keV  by synchrotron emission we take the parameters
$\Delta t^{obs}\simeq 1\, yr$, $\Gamma=4.2$, $ \xi_B=0.2$ and
$f_{es}=0.1$. The two parameters $\Gamma=4.2$ and $\Delta
t^{obs}\simeq 1$ yr
imply  that the jet in Cen A is mildly relativistic and the time
variability is order of year in this keV range of the photon
energy. Here we further assume that 
the same time variability also holds around 150 keV.
The parameter $ \xi_B=0.2$ signifies that the equipartition energy in
the magnetic field corresponds to 20\% of the total luminosity.
Using the above values of the parameters, the magnetic field is given
by,
\be
B= 1.8\times 10^{-4}\,\,\,
\sqrt{\frac{\xi_B}{0.2}}\biggl(\frac{\Gamma}{4.2}\biggr)^{-3}\biggl(\frac{\Delta
  t^{obs}}{1 yr}\biggr)^{-1}\biggl(\frac{L_\gamma^{obs}}{3.78\times
  10^{42} erg/s}\biggr)^{1/2}\, G.
\label{bfield}
\ee
Also the time variability corresponds to a distance scale of $r\sim 2 \Gamma^2
\Delta t^{obs}\simeq 10.8\, pc$ which is in the nuclear region of Cen 
A.

Cen A has a dusty torus within 100 pc of the black hole, with high
column density and there can be infrared emission from it.
Additionally we also assume that the first peak at around  $\sim 4
\times 10^{-4}$ eV in the nuclear SED has contribution from electron
synchrotron emission.  But for
electron synchrotron emission to be operative in this region 
the time variability is much smaller compared to the one for 150
keV peak and one can consider the minimum variability time scale $\Delta t^{obs}\sim 1$
day \cite{Kinzer:1995}.  To fix the first peak we  keep the bulk
Lorentz factor $\Gamma=4.2$ as before but vary $\xi_B$
and $f_{es}$. In this case both $\xi_B$ and $f_{es}$ are small
compared to their corresponding values for 150 keV . This shows
that for low energy peak the synchrotron cooling is very very slow and also the
equipartition energy in the magnetic field is very small. As discussed
in ref.\cite{Gupta:2007yb}, the energy dependence of the differential
photon spectrum for electron synchrotron emission is  proportional to $E_{\gamma}^{-2/3}$ below
the break energy  $\sim 4 \times 10^{-4}$ eV.  In a very recent paper
by Araya and Cui\cite{Araya:2010pq} it is argued that a diffuse
electron population that is uniformly
distributed over the entire supernova remnant of Cassiopia A  can
contribute to radio and infrared fluxes. Also in ref.\cite{Capetti:2000vh,Hardcastle:2006sp} it is shown that in the
infrared region of Cen A the best fitted photon spectral index lies
around $0.89\pm 0.25$ and $0.84\pm0.18$ which further
strengthen the non-thermal synchrotron origin of the infrared emission.

It has been suggested that both radio and X-ray emission in the inner
jet of Cen A are produced by relativistic electrons with Lorentz
factor $\gamma\simeq 8\times 10^7$ in 60
$\mu$G magnetic field\cite{Schreier:1981}.  With our fitted parameters 
we found that the magnetic field lies very close to 60 $\mu$G
as shown in Eq.(\ref{bfield}) and at the same time the
electron Lorentz factor at the break energy is $\gamma \simeq 2.3
\times 10^7$. The maximum  observed photon energy for 
the above Cen A parameters is given by
\be
E^{obs}_{\gamma, max}=\frac{9 \pi e^2}{8\sigma_T\eta m_e}
\frac{\Gamma}{\beta^2} \simeq 1.65\, GeV,
\label{egammaobsmax}
\ee
where $\eta\sim 1$.  If the second peak of nuclear SED is due to
electron synchrotron emission,  the maximum observed photon energy from
this process is 1.65 GeV and the  synchrotron photon spectrum is given by
\be
\left ( \frac{dN_{\gamma}}{dE_{\gamma}}\right )_{syn} = \frac{A_e}{2} \left ( \frac{1}{C_e E_0}\right )^{\frac{(-\alpha+2)}{2} }
 \left\{ 
\begin{array}{l l}
\left ( \frac{E_{\gamma, b}}{E_0} \right )^{-\frac{1}{2}}
 \left ( \frac{E_{\gamma}}{E_0} \right )^{-\frac{(\alpha+1)}{2}} & \quad
  E_{\gamma} < E_{\gamma, b} \\
  \left ( \frac{E_{\gamma}}{E_0}\right )^{-\frac{(\alpha+2)}{2}} & \quad E_{\gamma} \ge E_{\gamma,b}\\
\end{array} \right.,
\label{synspect}
\ee
where
\be
C_e = \frac{3\pi e}{8m^3_e} \frac{1}{\Gamma^4} \left (
  \frac{\xi_B}{2}\right )^{1/2} 
\frac{1}{\Delta t^{obs}} (L^{obs}_{\gamma})^{1/2},
\ee
and $E_{\gamma,b}=150$ keV is the break in photon energy corresponding to the
break in electron energy.  Inclusion of  the first break at
far-infrared energy in Eq.(\ref{synspect}), the condition $E_{\gamma}
< E_{\gamma, b} $ will be replaced by $4\times 10^{-4}\, eV <
E_{\gamma} < E_{\gamma, b} $.
During the period 1991-1995, the observation
of Cen A by Compton Gamma-Ray Observatory (CGRO)  revealed the break
in the spectral energy distribution $\nu F_{\nu}$ around 150 keV with
a maximum flux of about $\sim 10^{-9}\, erg/cm^2/s$\cite{Kinzer:1995}.  By considering
a mean power law index $\sim 1.85$  the observed
luminosity in the energy range 40 keV to 1200
keV is $L_{\gamma}^{obs}=3.78\times 10^{42} erg/s$\cite{Bond:1996}
where the differential photon spectrum can be given by
\be
\left ( \frac{dN_{\gamma}}{dE_{\gamma}}\right )^{obs}=A_{\gamma} \left (\frac{E_{\gamma}}{E_{0}}\right )^{-\lambda},
\label{ngamma}
\ee
with 
$A_{\gamma} \simeq 5.87 \times 10^{-5}/keV/cm^2/s$.
But it is observed that below 150 keV the power index   
$\lambda\sim 1.8-2.0$ and  for $E_{\gamma} \ge E_{\gamma,b}$ it is $2.4\pm 0.28 $ .
The observed spectrum is related to the synchrotron spectrum at the
source as 
\be
\left ( \frac{dN_{\gamma}}{dE_{\gamma}}\right )^{obs} \simeq 
\frac{4\Gamma^4 (\Delta t^{obs})^2}{d^2_z }  e^{-\tau_{int}} \left ( \frac{dN_{\gamma}}{dE_{\gamma}}\right )_{syn},
\label{obssynspct}
\ee
where $d_z=3.8\, Mpc$ is the distance of Cen A from us,  and $\tau_{int}$
is the optical depth of photon in the jet. The term $e^{-\tau_{int}} $
in the above equation corresponds to the decrease in observed spectrum due to
$\gamma\gamma$ interaction in the photon background of the jet. In
this energy range we have observed that $\tau_{int}\le 10^{-8}$ and
does not  reduce the observed photon
spectrum\cite{Bhattacharjee:2002it} which is shown in Figure 1 for
different observed luminosities. So for low energy, one
can neglect the contributions due to $\gamma\gamma$ interactions. Here
we use $\alpha=2.7$ as seen in the diffuse UHECR flux just before the
Greisen-Zatsepin-Kuzmin cut off assuming that like proton, electron
also follows the same power spectrum. So in Eq.(\ref{synspect}), 
$\alpha=2.7$ is consistent with the observed with the observed
photon spectrum as described  in Eq.(\ref{ngamma}).
Comparison of Eq.(\ref{ngamma}) with Eq.(\ref{obssynspct}) gives
$A_e \sim 5.4\times 10^{12}/keV/cm^2/s$.

\begin{figure}[htb]
  \begin{center}
    \includegraphics[height=10.0cm]{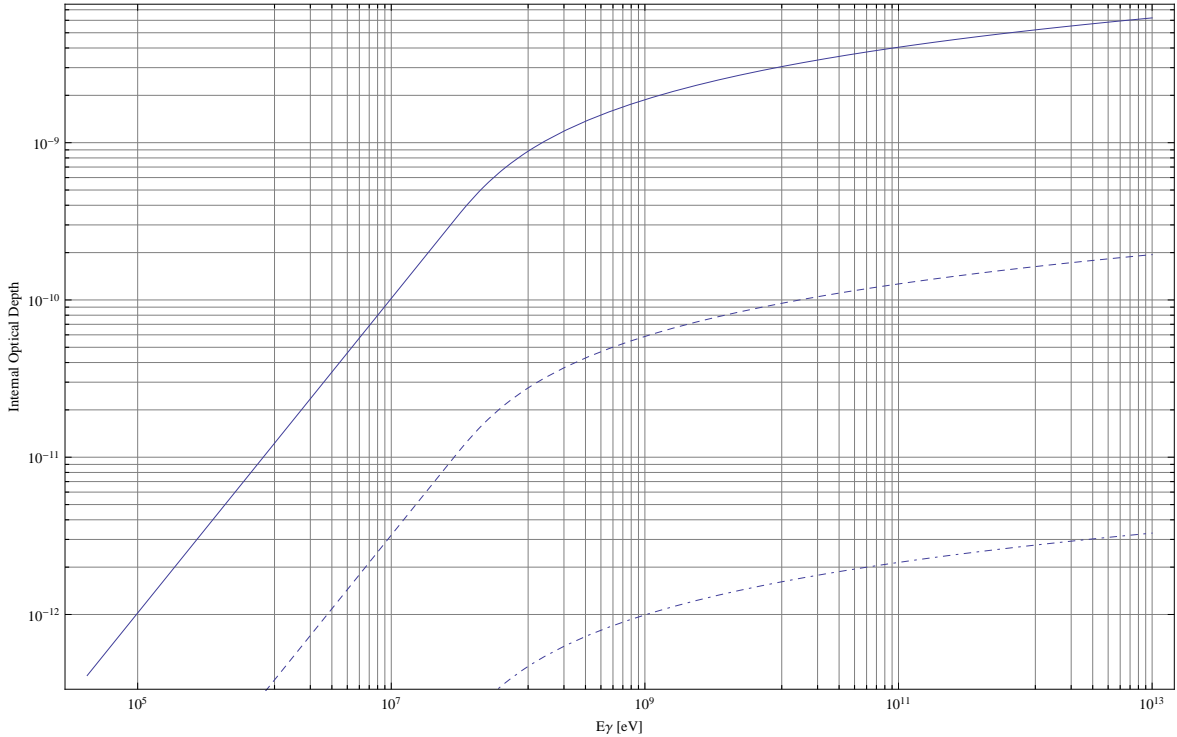}
  \end{center}
  \caption{We have shown the internal photon optical depth $\tau_{\gamma\gamma}$ as a
    function of photon energy $E_{\gamma}$ in the nuclear region of
    the jet for three different luminosities. These are (a) continuous line is
    for $L^{obs}_{\gamma}=3.78 \times 10^{42}\, erg/s$, (b) Dashed line is
    for $L^{obs}_{\gamma}=1.18 \times 10^{41}\, erg/s$ and (c) Dot-dashed
    line is for $L^{obs}_{\gamma}=2.0 \times 10^{39}\, erg/s$.}
\end{figure}

In the Inverse Compton (IC)  scattering process,  the energy of an IC
photon is very high and for $40\, keV$ minimum observed photon energy
the electron Lorentz factor is $\gamma\sim 6.2\times 10^7$.  So for IC to be operative in
this region, the photon energy has to be really very small and for
this reason we do not consider the contribution from IC
process\cite{Israel:1998ws, Feigelson:1981}. The combined 
OSSE (0.05 - 4 MeV) \cite{Kinzer:1995}, COMPTEL\cite{Steinle:1998}
 and EGRET (0.1 - 1GeV) \cite{Sreekumar:1999xw}
 data is fitted by
doubly broken power law with two break energies ($\sim$ 150 keV and
$16.7^{+27.3}_{-16.3}$ MeV). The spectral indices are $\lambda \simeq
1.74$ for $E_{\gamma} < 150$ keV, $\lambda \simeq
2.3 $ for $150 keV \le  E_{\gamma} <  16.7^{+27.3}_{-16.3}$ MeV,
otherwise $\lambda\simeq 3.3$ for $E_{\gamma}  >16.7^{+27.3}_{-16.3} $
MeV.  But the second break at $E_{\gamma}  >16.7^{+27.3}_{-16.3} $ has
large uncertainty and also the spectrum is almost smooth around and beyond this
energy. So in the present analysis we do not take into account for
this break and assume that the power index is 2.35 well fitted.

We have shown  in Eq.(\ref{egammaobsmax}) that the electron
synchrotron process contributes up to maximum observed photon energy
$\sim 1.65\; GeV$. Also in the nuclear  region $r\sim 10.8\, pc$, the
inelastic collision of Fermi accelerated protons with the background
protons will produce $\gamma$-rays due to decay of neutral
pions\cite{Halzen:2008vz}. 
In this case the minimum observed energy of proton is  $E^{obs}_p \simeq
5. 2\,GeV$ of which about $18\%$ goes to a 
photon\cite{Cavasinni:2006nx,Crocker:2004bb} which corresponds to
$E^{obs}_{\gamma} \simeq 0.93\, GeV$. So for $E^{obs}_{\gamma} \ge
0.93\, GeV$, $pp$ collision will also produce $\gamma$-rays and the
projectile high energy protons (due to Fermi acceleration) do not have a
break energy (the only break in proton energy will come from the proton
synchrotron emission which is given by $E_{p,b} \simeq 2.8\times
10^{27}\, eV$).  In the nuclear region the same  Fermi accelerated
protons colliding with the background photon should also produce
photons through delta resonance and its flux  can also compete with
the one from $pp$
process. By considering the highest detected energy bin found by the
Energetic Gamma-Ray Experiment telescope (EGRET) at a break energy
$\simeq 200$ MeV \cite{Sreekumar:1999xw, :2007qd}, 
we observed that the $p\gamma$ process
starts to contribute at energy $E_{\gamma,min}\simeq 1.4$ GeV. But
the flux ratio of $pp$ to $p\gamma$ is about $\sim 3.13\times 10^{14}$
at 1.4 GeV photon energy which increases by a very small amount for
higher energies which shows that $p\gamma$ process is
negligible and does contribute to the very high energy gamma rays.

In the observed photon energy range 0.93 GeV to 1.65
GeV both electron synchrotron and $pp$ interaction will
contribute and the $pp$ contribution can be given by
\be
\left ( \frac{dN_{\gamma}}{dE_{\gamma}}\right )_{\pi^0 pp} = 
f_{\pi0 pp} \left ( \frac{2}{\delta}\right )^{2-\alpha}  \frac{A_p}{\Gamma^{-\alpha}} 
\left ( \frac{E_{\gamma}}{E_0} \right )^{-\alpha},
\ee
where
\be
f_{\pi0 pp} =\Gamma\Delta t^{obs} n_{p} \sigma_{pp},
\ee
$\delta=0.18$ and $n_{p}$  is the proton density in the
jet and $\sigma_{pp}\simeq 4.5\times 10^{-26}\, cm^2$. The energy normalization we have used here is 
$E_0=1\, GeV$.  Around $r\sim 10.8$ pc from the central black hole the proton density  can be in the
range of $10^4/cm^3$ to $10^6/cm^3$ ref.\cite{Halzen:2008vz,Israel:1998ws}. Here we consider an
intermediate value of $n_p\sim 10^5/cm^3$ and  the optical depth of the process is $\tau_{pp}=rn_p
\sigma_{pp} \simeq 0.15$, which shows that about 15\% of the protons will
interact to produce photons. In this energy range the photon power
spectrum from both electron synchrotron and $pp$ 
are proportional to $E_{\gamma}^{-(\alpha+2)/2}$  and
$E_{\gamma}^{-\alpha}$ respectively. Also it is important to note that
we have not observed any break in photon energy around 0.93 GeV and
beyond and on
top of that the photon power spectrum is well fitted with the index
$2.4\pm 0.28$ in the energy range 30 MeV to 10 GeV
\cite{Sreekumar:1999xw}.  
The EGRET has measured the average flux above 100 MeV  (100
MeV to 10 GeV) from Cen A which is $N_{\gamma}=(13.6\pm 2.5) \times
10^{-8}\, photons/cm^2/s$ and for $\lambda=2.35$  
the luminosity obtained is $L_{\gamma}^{obs}=1.19\times 10^{41}\,
erg/s$. As we have discussed earlier, from 100 MeV to 0.93 GeV, the
luminosity is only due to the electron synchrotron radiation which
is given as $L_{\gamma}\simeq 9 . 24\times 10^{40}\, erg/s$. On  the
other hand in the energy range 0.93 GeV to 1.65 GeV, the luminosity
from electron synchrotron is $L_{\gamma}\simeq 1.85\times 10^{40}\,
erg/s$ and from $pp$ is $L_{\gamma}\simeq 3.23\times 10^{39}\,
erg/s$. Finally the luminosity in the range 1.65 GeV to 10 GeV from
$pp$ process is $L_{\gamma}\simeq 4 . 69\times 10^{39}\, erg/s$.
So we have to adjust the parameters $n_{p}$ and $A_p$ so
that photon spectrum from $pp$ interaction will overlap with the one
from electron synchrotron. But once we fix the value of the parameter $n_p$ in the
nuclear region (we take $n_{p}\simeq 10^5/cm^3) $, only $A_p$ has
to be adjusted. The observed photon spectrum in the energy range 0.93
GeV to 1.65 GeV is
\be
\left ( \frac{dN_{\gamma}}{dE_{\gamma}}\right )^{obs} \simeq 
\frac{4\Gamma^4 (\Delta t^{obs})^2}{d^2_z }  e^{-\tau_{int}} 
\left [
\left ( \frac{dN_{\gamma}}{dE_{\gamma}}\right )_{syn} + 
\left ( \frac{dN_{\gamma}}{dE_{\gamma}}\right )_{\pi^0 pp} \right ].
\ee
The fitted value of $A_p$ is given by $A_p=1.8\times
10^3/GeV/cm^2/s$.  Now we shall turn to the high energy range
i.e. above 250 GeV in which many collaborations have measured the
photon flux and also the spectral index. As we have discussed
$p\gamma$ process is negligible compared to the $pp$ process, so only
$pp$ process will contribute to the very high energy gamma-ray
production in Cen A. For $E_{\gamma} \ge 250$ GeV only $pp$ process
will contribute.  But before that we would like
to estimate the  number of events from PAO which is shown in the next section. 

\section{Cosmic Ray Events from Pierre Auger Observatory}

The Pierre Auger Observatory (PAO) had reported a correlation between
ultra high energy cosmic rays (UHECR) and nearby AGN within $\sim$75
Mpc\cite{Abraham:2007si,Lemoine:2009zza}. 
Roughly 10 cosmic ray events are concentrated around the Centaurus
direction, a region with a high density of AGNs and two of these events 
fall within 3 degrees  from Cen A suggesting that it could be the
first evidence of the UHECR source.  Although the composition of these
CR events are  unknown, we assume them to be protons\cite{Kachelriess:2008qx,Gupta:2008tm}.

In general the intergalactic magnetic field plays a major role in
deflecting the CR protons from their original direction. But the
deflection angle $\Delta\theta$ is inversely proprotional to the CR proton energy and
for a 57 EeV proton it is less than a degree, if we take the
intergalactic magnetic field to be of order $10^{-8}$ G and for higher
energy CR protons, the deflection will be still smaller. So the
propagation of CR protons from Cen A with energy $\ge 57$ EeV are
almost ballistic and one can neglect the effect of magnetic field on it.

We estimate the expected number of events  from Auger
array. For a point source, the integrated exposure of PAO  is
$\Xi= 9000/\pi km^2$ yr and the integral flux of UHECR from
PAO can be expressed as
\be
N(>E_{min}) = \frac{\#\, Events}{Exposure},
\ee
where $E_{min}=57$ EeV.
One has to also consider the relative exposure  $\omega
(\delta)$ for angle of declination $\delta$. For Cen A, $\delta =
47^{\circ}$ and the corresponding value of $\omega
(\delta)\simeq 0.64$\cite{:2007qd}.  The time duration for the
collection of data by PAO is about 15/4 yr between 1st January 2004
and August 2007. So from the above equation we obtain
\be
\#\, Events =  \frac{9\times 10^3\, km^2\, yr} {\pi}\,  N(>E_{min})\, \omega
(\delta) \, \frac{15}{4} e^{-\tau_{pp}},
\label{nevents}
\ee
where $e^{-\tau_{pp}}$ is the survival probability of proton due to
interaction with the protons in the jet background. Also the integral
flux is related to the proton normalization constant $A_p$ as
\be
N(>E_{min}) = \frac{A_p}{\alpha-1} \left ( \frac{E_{min}}{E_0}
\right )^{-\alpha} E_{min}.
\label{intgflux}
\ee
Putting Eq.(\ref{intgflux}) in Eq.(\ref{nevents}) and simplifying we
obtain
\be
\#\, Events =5.75\times 10^2 \times A_{p0},\ee
where we have defined $A_p=A_{p0}\, GeV^{-1}\,cm^{-2}\,s^{-1}$.
We can obtain the value of $A_{p0}$ from different observations/Experiments
which are discussed in the next section and using these values we can
calculate the expected number of events in these observations.

\section{Observation of Cen A above 250 GeV}

As discussed in the introduction, Cen A has been observed in almost
all wavelengths. Following are the observations of Cen A in Very High Energy
$\gamma$-rays ($\ge$ 250 GeV)
\cite{Clay:1994uy,Clay:2010id,Allen:1993ir}.
From each observations we have used their given integral flux 
and calculate the photon normalization constant $A_{\gamma}$ as given
in Eq.(\ref{ngamma}) and after that comparing with the observed
spectrum of $pp$ process calculate $A_p$.

\begin{enumerate}
\item
The H.E.S.S. Collaboration operates an array of four large imaging
Cherenkov telescopes to detect Very High Energy (VHE) $\gamma$-rays in
located in Southern hemisphere in Namibia. Cen A was observed between
April 2004 and July 2008 with total live time of 115.0 hours.  The measured integral
flux above 250 GeV energy threshold was $(1.56\pm 0.67)\times 10^{-12}
\, cm^{-2}\, s^{-1} $ which was around the central region of the
galaxy.  Also the measured differential photon spectrum
is well described by a power-law with the index $\sim 2.73$ and the
normalization constant 
$A_{\gamma}\sim 2.45\times
10^{-13}\,/TeV/cm^2/s$\cite{Aharonian:2009xn}. 
This gives the luminosity
$L_{\gamma}\sim 2.58\times 10^{39}\,erg/s$ and $A_{p0}=0.18 \times
10^{-3}$.\\
\item
In March 1997, with the Durham Mk6 telescope, Cen A was observed for
6.75 hours with a 3$\sigma$ flux upper limit above 300 GeV of
$5.2\times 10^{-11}\, cm^{-2}\, s^{-1}$\cite{Chadwick:1999my}. This
gives $A_{\gamma}\sim 1.14\times 10^{-11}\,/TeV/cm^2/s$.\\
\item
The JANZOS group observed Cen A from New Zealand during April 1988 and
June 1989 for 56.9 hours and reported a 95\% confidence level upper 
limit on the flux above 1 TeV of $3.74\times 10^{-11}\, cm^{-2}\,
s^{-1}$ \cite{Allen:1993ep} which gives $A_{\gamma}\sim 1.14\times 10^{-11}\,/TeV/cm^2/s$.\\
\item
In March and April 1999, Cen A was observed by the CONGAROO 3.8 m
telescope for 45 hours with a   3$\sigma$ flux upper limit above 1.5 TeV of
$5.5\times 10^{-12}\, cm^{-2}\, s^{-1}$ for a point source at the core
of the galaxy  \cite{Rowell:1999fx} and this gave $A_{\gamma}\sim 1.86\times 10^{-11}\,/TeV/cm^2/s$.\\
\item
In March and April of 2004 further observations were made to Cen A
with three 10 m telescopes of the CONGAROO-III array for 10.6 hours.
For the core region of Cen A with a 2$\sigma$ flux upper limit above
424 GeV was $4.9\times 10^{-12}\, cm^{-2}\, s^{-1}$
\cite{Kabuki:2007am} and we obtain 
$A_{\gamma}\sim 1.94\times 10^{-12}\,/TeV/cm^2/s$.\\

\end{enumerate}

In all of the above VHE $\gamma$-ray observations we have fitted the  differential
photon spectrum with the power index 
$\alpha=2.7$\cite{Aharonian:2009xn,Raue:2009vp,Grindlay:1975} and calculated their
normalization constant $A_{\gamma}$. Comparison of $A_{\gamma}$ with
the observed photon spectrum from the respective process gives the
values of $A_p$ which are given in table 1. Using this $A_p$ we
calculate the number of events from these different observations.

It is observed that the H.E.S.S. collaboration results give less than
one event during the observation period of 15/4 years. On the other
hand during the same observation period the Durham Mk6 telescope,
CANGAROO-III, JANZOS and CANGAROO observations could have given
respectively 5.06, 0.86, 16.61 and 8.26 events in PAO due to $pp$
interaction in Cen A.  So our analysis shows that many UHECR protons
from Cen A can be observed by PAO.

\section{Conclusions}

The Cen A is observed in almost all the energy band and its nuclear
SED shows two peaks of which one is around 150 keV and another is in
the far-infrared band. In this work we assumed that the peak around
150 keV is due to the electron synchrotron radiation and  adjust the
parameters $\Gamma$, $\xi_B$ , $\Delta t^{obs}$
and $f_{es}$  so that the observed photon energy $E_{\gamma}^{obs}$ is
150 keV. In this fitting we have $\Gamma=4.2$ and $\Delta t^{obs}=1$
yr correspond to the fact that the jet is in the nuclear region $r\sim
10.8$ pc and it is mildly relativistic which is being observed in Cen
A.  The magnetic field obtained around the nuclear region is about 180
$\mu\,G$\cite{Schreier:1981}.  We found that the IC process does not
contribute to the observed photons in the GeV-TeV range. In our model
the maximum energy up to which the electron synchrotron contributes is
1.65 GeV and the other process which
contributes to high energy (1 to 10 GeV) to very high energy ($\ge$
250 GeV)  is the $pp$ process. Unfortunately the $p\gamma$
process is very small compared to the $pp$ process. 

We compared the observed photon spectrum from different observations
by taking the spectral index $\alpha=2.7$ with the calculated one from
the $pp$ process, which determines the normalization constant $A_p$ for the
cosmic ray (CR) protons. We use the $A_p$ from these observations to
calculate the number of events from PAO and found that indeed many CR
protons above 57 EeV can arrive on Earth from Cen A.  In a recent
paper in the context of Cassiopia A, a young supernova remnant, Araya
and Cui\cite{Araya:2010pq} have also independently come to the same
conclusion that synchrotron emission from the electron can account for
the radio to X-ray emissions and additional hadronic component is
needed to explain the GeV emission thus providing evidence for the
production of cosmic rays.

\begin{table}
\begin{center}
\caption{Observation of Cen A by different Experiments with their
  respective lower observed photon energy limits are shown here. We
  also show their corresponding proton normalization constant and the
  expected number of CR proton events.\\}
\renewcommand{\tabcolsep}{0.35cm}
\renewcommand{\arraystretch}{1.05}
\begin{tabular}{|c|c|c|c|}\hline
 Experiment [ref.] & $E_{\gamma} $  &$A_{p0}$&$\#$\,  Events  \\ \hline 
 H.E.S.S.\cite{Aharonian:2009xn}   & $\ge \,250$ GeV    &  $0.18\times
 10^{-3}$  &0.1   \\ \hline
 Durham\cite{Chadwick:1999my}      & $\ge \,300$ GeV    &  $8.81\times
 10^{-3}$  &5.06  \\ \hline
 CANGAROO-III \cite{Kabuki:2007am} & $\ge \,424$ GeV    &  $1.5\times 10^{-3}  $  &0.86  \\ \hline
 JANZOS \cite{Allen:1993ep}        & $\ge \,\,\,\,\,\,1$ TeV      &
 $28.9 \times 10^{-3}$  &16.61 \\ \hline
 CANGAROO \cite{Rowell:1999fx}     & $\ge \,\,\,\,1.5$TeV     &
 $14.38\times 10^{-3}$  &8.26  \\ \hline
\end{tabular}
\label{aa}
\end{center}
\end{table}


We are grateful to  S. Nagataki and Y. Y. Keum for many useful discussions. 
This work is partially supported by DGAPA-UNAM (Mexico) Project 
No. IN101409  and Conacyt Project No. 103520 (SS) and by 
NASA NNX09AO94G and NSF AST-0908362 (BZ).


\end{document}